\documentclass[11pt,epsf]{article}
\usepackage{epsfig}
\newcommand{\nc}{\newcommand}
\nc{\postscript}[2] 
{\setlength{\epsfxsize}{#2\hsize}\centerline{\epsfbox{#1}}}
\nc{\non}{\nonumber}
\nc{\pt}{p_{{}_T}}
\nc{\lmc}{\Lambda_c^+}
\nc{\plmc}{\vec{\Lambda}_c^+}
\nc{\all}{A_{LL}}
\nc{\dg}{\Delta G(x,Q^2)}
\nc{\stil}{\tilde{s}}
\nc{\ttil}{\tilde{t}}
\nc{\util}{\tilde{u}}
\nc{\shs}{\hat{s}}
\nc{\ths}{\hat{t}_1}
\nc{\uhs}{\hat{u}_1}
\nc{\cosec}{\rm cosec}
\nc{\mpr}{m_p}
\nc{\mc}{m_c}
\nc{\mlc}{m_{\lmc}}
\nc{\pa}{p_{{}_a}}
\nc{\pb}{p_{{}_b}}
\nc{\pA}{p_{{}_A}}
\nc{\pB}{p_{{}_B}}
\nc{\plc}{p_{\lmc}}
\nc{\pc}{p_{{}_c}}

\def\dps{\displaystyle}
\def\mib#1{\mbox{\boldmath $#1$}}

\nc{\prd}[3]{{\it Phys.\ Rev.}\ {{\bf D{#1}} (#2), #3}}
\nc{\prl}[3]{{\it Phys.\ Rev.\ Lett.}\ {{\bf {#1}} (#2), #3}}
\nc{\plb}[3]{{\it Phys.\ Lett.}\ {{\bf B{#1}} (#2), #3}}
\nc{\npb}[3]{{\it Nucl.\ Phys.}\ {{\bf B{#1}} (#2), #3}}
\nc{\ptp}[3]{{\it Prog.\ Theor.\ Phys.}\ {{\bf {#1}} (#2), #3}}
\nc{\zfp}[3]{{\it Z.\ Phys.}\ {{\bf C{#1}} (#2), #3}}
\nc{\mpla}[3]{{\it Mod.\ Phys.\ Lett.}\ {{\bf A{#1}} (#2), #3}}
\nc{\rmp}[3]{{\it Rev.\ Mod.\ Phys.}\ {{\bf {#1}} (#2), #3}}
\nc{\ijmpa}[3]{{\it Int.\ J.\ of\ Mod.\ Phys.}\
               {{\bf A{#1}} (#2), #3}}
\nc{\epj}[3]{{\it Eur.\ Phys.\ J.}\ {{\bf C{#1}}  (#2), #3}}
\begin{document}
\pagestyle{empty} \setlength{\footskip}{2.0cm}
\setlength{\oddsidemargin}{0.5cm} \setlength{\evensidemargin}{0.5cm}
\renewcommand{\thepage}{-- \arabic{page} --}
\def\mib#1{\mbox{\boldmath $#1$}}
\def\bra#1{\langle #1 |}      \def\ket#1{|#1\rangle}
\def\vev#1{\langle #1\rangle} \def\dps{\displaystyle}
%
   \def\thebibliography#1{\centerline{\large \bf REFERENCES}
     \list{[\arabic{enumi}]}{\settowidth\labelwidth{[#1]}\leftmargin
     \labelwidth\advance\leftmargin\labelsep\usecounter{enumi}}
     \def\newblock{\hskip .11em plus .33em minus -.07em}\sloppy
     \clubpenalty4000\widowpenalty4000\sfcode`\.=1000\relax}\let
     \endthebibliography=\endlist
   \def\sec#1{\addtocounter{section}{1}\section*{\hspace*{-0.72cm}
     \normalsize\bf\arabic{section}.$\;$#1}\vspace*{-0.3cm}}
\vspace*{-1.8cm}
\begin{flushright}
$\vcenter{
\hbox{KOBE-FHD-02-01}
\hbox{(hep-ph/0201144)}
}$
\end{flushright}
\renewcommand{\thefootnote}{$\dag$}
\vskip 0.2cm
\begin{center}
{\Huge \bf Charmed Hadron Production \\
\vskip 0.5cm
at RHIC}\footnote{
Talk presented by K. OHKUMA at 3rd Circum-Pan-Pacific Symposium on 
``High Energy Spin \\
\phantom{abc} Physics'', Beijing, China,Oct. 8-13, 2001 
(talk on Oct. 11, 2001)
}

\vskip 0.15cm
\end{center}

\vspace*{0.4cm}
\begin{center}
\renewcommand{\thefootnote}{\alph{footnote})}
 \setcounter{footnote}{0}
{\Large
{\sc Kazumasa OHKUMA$^{\:1),\:}$}\footnote{E-mail address:
\tt ohkuma@radix.h.kobe-u.ac.jp},\ \ \ \ 
{\sc Toshiyuki MORII$^{\:1,2),\:}$}\footnote{E-mail address:
\tt morii@kobe-u.ac.jp}\\
\vskip 0.5cm and \\
\vskip 0.5cm 
{\sc Satoshi OYAMA$^{\:1),\:}$}\footnote{E-mail address:
\tt satoshi@radix.h.kobe-u.ac.jp}\,
}
\end{center}
{\large
\vspace*{0.4cm}
\vskip 0.2cm
\centerline{\sl $1)$ Graduate School of Science and Technology,
Kobe University}
\centerline{\sl Nada, Kobe 657-8501, JAPAN}

\vskip 0.2cm
\centerline{\sl $2)$ Faculty of Human Development, 
Kobe University}
\centerline{\sl Nada, Kobe 657-8501, JAPAN}
}
\vspace*{1.8cm}
\centerline{\Large ABSTRACT}

\vspace*{0.2cm}
\baselineskip=23pt plus 0.1pt minus 0.1pt
{\Large
To extract information about polarized gluon distribution in the proton, 
charmed hadron, actually $\Lambda_c^+$,
productions at RHIC experiment are studied.
We found that the spin correlation asymmetry between the initial proton and 
the produced $\Lambda_c^+$ is enable us to distinguish parameterization
models of polarized gluons. \\
\vfill
}

\newpage
\renewcommand{\thefootnote}{$\sharp$\arabic{footnote}}
\pagestyle{plain} \setcounter{footnote}{0}
\baselineskip=21.0pt plus 0.2pt minus 0.1pt

\section{Introduction}	
Since the surprising EMC measurement$^1$ of
the polarized structure function of proton $g_1^p(x,Q^2)$ was reported 
more than ten-years ago,
the spin structure of the proton is  still mysterious problem being 
called the  proton spin puzzle$^2$. 
As is well known, a proton is not an elementary but compound particle
and thus, its spin is carried by its constituents as described by a 
sum rule,
\begin{equation}
\frac{1}{2}=\frac{1}{2}\Delta \Sigma + \Delta G + <L_z>_{q+g}
\end{equation}
where $\frac{1}{2}$ on the left side means a spin of the proton, 
while $\Delta \Sigma, \Delta G$ and $<L_z>_{q+g}$ represent the amount of
the proton spin carried by the constituent quarks, gluons and 
their orbital angular momenta, respectively.
An extensive study on polarized structure functions of
nucleons brought about a rather good information of
valence quark distributions in the proton,
However,
knowledge of $\Delta G$ and $<L_z>_{q+g}$ is 
still poor because it is very difficult to directly extract its
information from existing experimental data. 
In this work, we are interesting in the polarized gluon distribution
$\Delta G$.
So far, to extract its information, 
many processes depending on the gluon interactions
have been proposed and studied.

Here, to study the polarized gluon distribution in a proton,
we propose another process,
 $$p\vec{p}\to \vec{\Lambda}_c^+X,$$
which could be observed in forthcoming BNL-RHIC experiment.
In this process, $\lmc$ is dominantly produced via fragmentation 
of a charm quark originated from gluon--gluon fusion
\footnote{Since 
charm quarks are tiny contents in  the proton, 
the gluon--gluon fusion process is dominant for charm quark production.}.
Thus, its cross section is directly proportional to the 
gluon distribution in the proton. 
Moreover, the $\lmc$ is composed of a heavy quark $c$ and antisymmetrically
combined light $u$ and $d$ quarks.  
Hence, the $\lmc$ spin is basically carried by a charm quark
through gluon--gluon fusion.
Therefore, observation of the spin of the produced
$\lmc$ gives us information about the polarized gluons in the proton\\
\vspace*{0.6cm}
\section{Spin Correlation Asymmetry}
To get information of $\Delta G$, we introduce the spin correlation
asymmetry of the target proton and produced $\Lambda_c^+$ baryon,$^3$
\begin{eqnarray}
A_{LL}&=&\frac{d \sigma_{++} - d\sigma_{+-} + d \sigma_{--}- d\sigma_{-+}}
{d \sigma_{++} + d \sigma_{+-}+d \sigma_{--} + d \sigma_{-+}}\nonumber\\
&\equiv&\frac{{d \Delta \sigma}/{d X}}
{{{d\sigma}/{d X}}},~~(X=\eta~{\rm or}~\pt), \label{eq:all}
\end{eqnarray}
where $d \sigma_{+ -}$, for example, denotes the spin-dependent
differential cross section with the positive helicity of the target proton
and the negative helicity of the produced $\Lambda_c^+$ baryon.
$\eta$ and $\pt$ mean pseudo-rapidity and transverse momentum of
produced $\lmc$, respectively. 
Moreover, according to quark-parton model,
$d \Delta \sigma / d X$ can be expressed as
\begin{eqnarray}
&&\frac{d \Delta \sigma}{d X}
 = \int^{Y{{}^{\rm max}}}_{Y{{}^{\rm min}}}
\int^{1}_{x^{{}^{\rm min}}_{{}_a}} 
\int^{1}_{x^{{}^{\rm min}}_{{}_b}}
          G_{p_{{}_A}\rightarrow g_{{}_a}}(x_a,Q^2)
   \Delta G_{\vec{p}_{{}_B} \rightarrow \vec{g}_{{}_b}}(x_b,Q^2)
   \Delta {D}_{\vec{c}\rightarrow \vec{\Lambda}_c^+}(z)\non \\
&&\phantom{\frac{d \Delta \sigma (\lambda,h)}{d \pt}=}
\times \frac{d \Delta \hat{\sigma}}{d \hat{t}}
J
dx_a dx_b dY,~~~\left(X,Y = \eta~{\rm or}~\pt ~(X \neq Y)\right),
\label{dcross}
\end{eqnarray}
where 
$ G_{p_{{}_A} \rightarrow g_a}(x_a,Q^2)$,
$\Delta G_{\vec{p}_{{}_B} \rightarrow \vec{g}_b}(x_b,Q^2)$
and $\Delta { D}_{\vec{c} \rightarrow \vec{\Lambda}_c^+}(z)$  
represent the unpolarized gluon distribution function, the polarized gluon
distribution function and the spin-dependent fragmentation function of the
outgoing charm quark decaying into a polarized $\plmc$, respectively.
 $d \Delta \hat{\sigma}/d \hat{t}$ is the spin correlation
 differential cross section in the subprocess 
and  $J$ is the Jacobian which transforms the variables  $z$  and $\hat{t}$ 
into  $\eta$ and $\pt$.  \\
\vspace*{0.6cm}
\section{Numerical Calculation}
To numerically estimate $A_{LL}$, 
we use, as input parameters,
$m_c = 1.25$ GeV, $m_p = 0.938$ GeV and $\mlc = 2.28$ GeV.$^4$
We limit the integration region of $\eta$ and 
$\pt$ of produced $\lmc$  as
$-1.3 \leq \eta \leq 1.3$ and
3 GeV $\leq \pt \leq$ 15(40) GeV for  $\sqrt{s}=200(500)$ GeV 
in order to get rid of  the contribution of
the diffractive $\lmc$ production and also the $\lmc$ production 
through a single charm quark production via $W$ boson exchange 
and $W$ boson production. 
In addition, 
we take the GRSV01$^{5}$ and AAC$^{6}$ parameterization models for the
polarized gluon distribution function and GRV98$^{7}$ for the
unpolarized one.
Though both of GSRSV01 and AAC models excellently reproduce the experimental 
data on the polarized structure function of nucleons $g_1(x)$,
the polarized gluon distributions for those models are quite different. 
In other words, the data on polarized structure function of nucleons 
$g_1(x)$ alone are not enough to distinguish the model of gluon distributions.
Since the process is semi-inclusive, the fragmentation function of a 
charm quark to $\Lambda_c^+$ is necessary to do numerical calculation.
For the unpolarized fragmentation function, 
we use Peterson fragmentation function$^{8}$, 
$D_{c \to \Lambda_c^+}(z)$.
However, since we have no data, at present, 
about polarized fragmentation
function for the polarized $\Lambda_c^+$ production, we take the
following ansatz for the polarized fragmentation function
$\Delta D_{\vec{c}\to \vec{\Lambda}_c^+}(x)$,
$$
\Delta D_{\vec{c}\to \vec{\Lambda}_c^+}(z)= C_{c\to \Lambda_c^+} 
D_{c \to \Lambda_c^+},
$$
where $C_{c \to \Lambda_c^+}$ is scale-independent spin transfer coefficient.
In this analysis, we study two cases: (A) 
$C_{c\to \Lambda_c^+}=1$ (non-relativistic quark model) and 
(B) $C_{c\to \Lambda_c^+}=z$ (Jet fragmentation model$^{9}$).

Numerical results of $A_{LL}$ are shown in Fig. 1 and Fig. 2.
As  shown in Fig. 1 and Fig. 2,  $A_{LL}$ is rather sensitive to the model of 
the polarized gluon distribution functions. 
Therefore, the process discussed here
could provide good information about the distribution of the polarized 
gluons in a nucleon. 
Especially,  the $\eta$ dependence of $A_{LL}$  at 
$\sqrt{s}$=200GeV is the  most effective to distinguish 
the parameterization models of polarized gluon 
because the magnitude of a numerical value of $A_{LL}$ is larger than
others.\\
\vspace*{0.6cm}
\section{Summary}
To extract information of polarized gluon distribution in the 
proton, we have proposed an interesting process;
$p\vec{p}\to\vec{\Lambda}_c^+ X$ and  
calculated the spin correlation asymmetry, $A_{LL}$, defined
by  Eq.(\ref{eq:all}). 
We found that in this process,
$A_{LL}$ is rather sensitive to the parameterization models of 
polarized gluon, and thus, process is quite promising for testing 
the models of polarized gluon distribution.
Error  estimation is important and 
now is undergoing.
To get better knowledge of $\Delta G$, we need more detailed 
information about the spin-dependent fragmentation function 
of a charm quark to $\Lambda_c^+$.\\
\vspace*{0.6cm}

\begin{figure}[htbn]
\hspace*{-1cm}
\parbox[b]{0.6\textwidth}
{
  \begin{center}
    \epsfxsize=7.0cm
\epsfbox{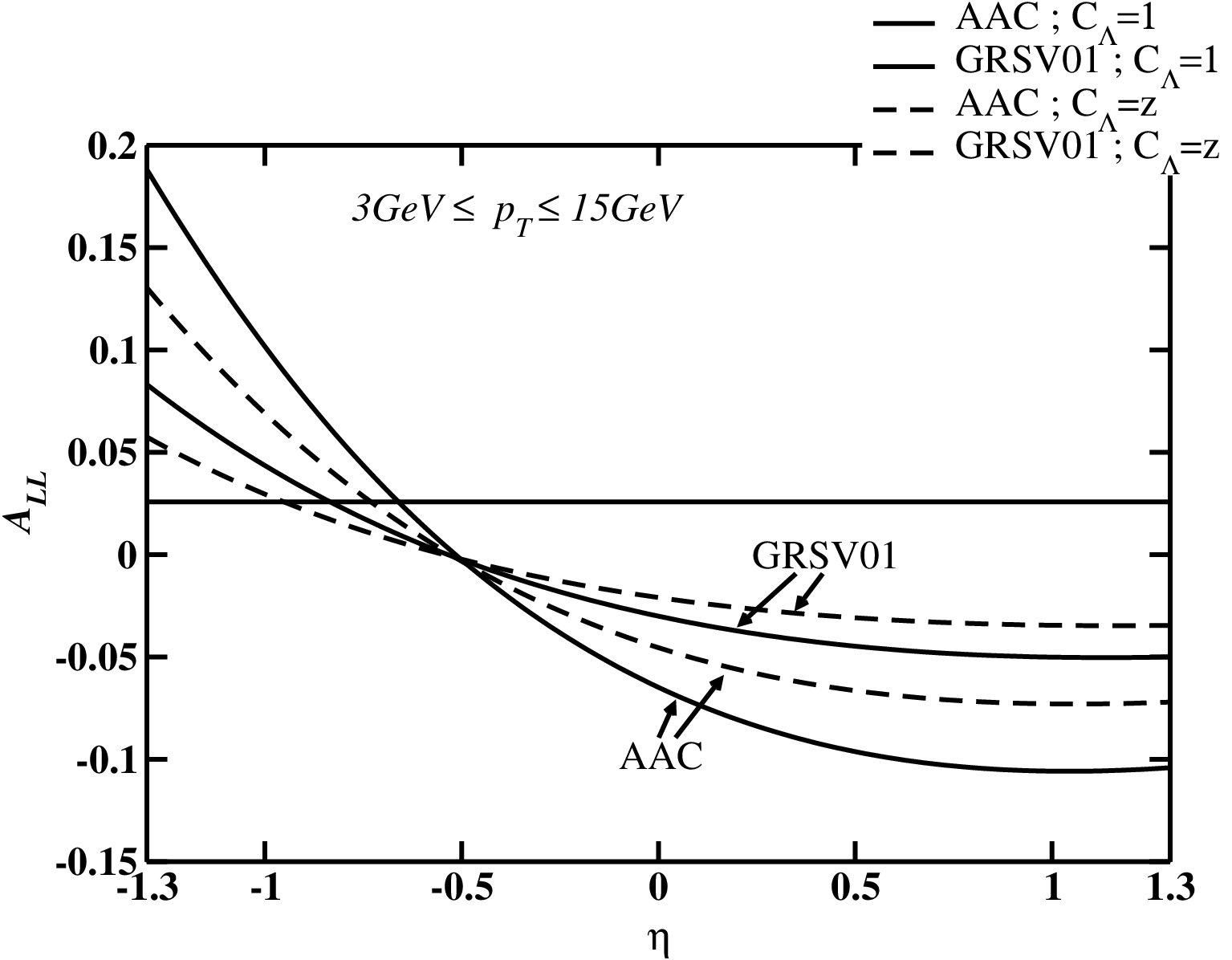}
  \end{center}
  \vspace*{-0.5cm}
} \hfill
\parbox[b]{0.5\textwidth}
{
  \begin{center}
    \epsfxsize=7.0cm
 \epsfbox{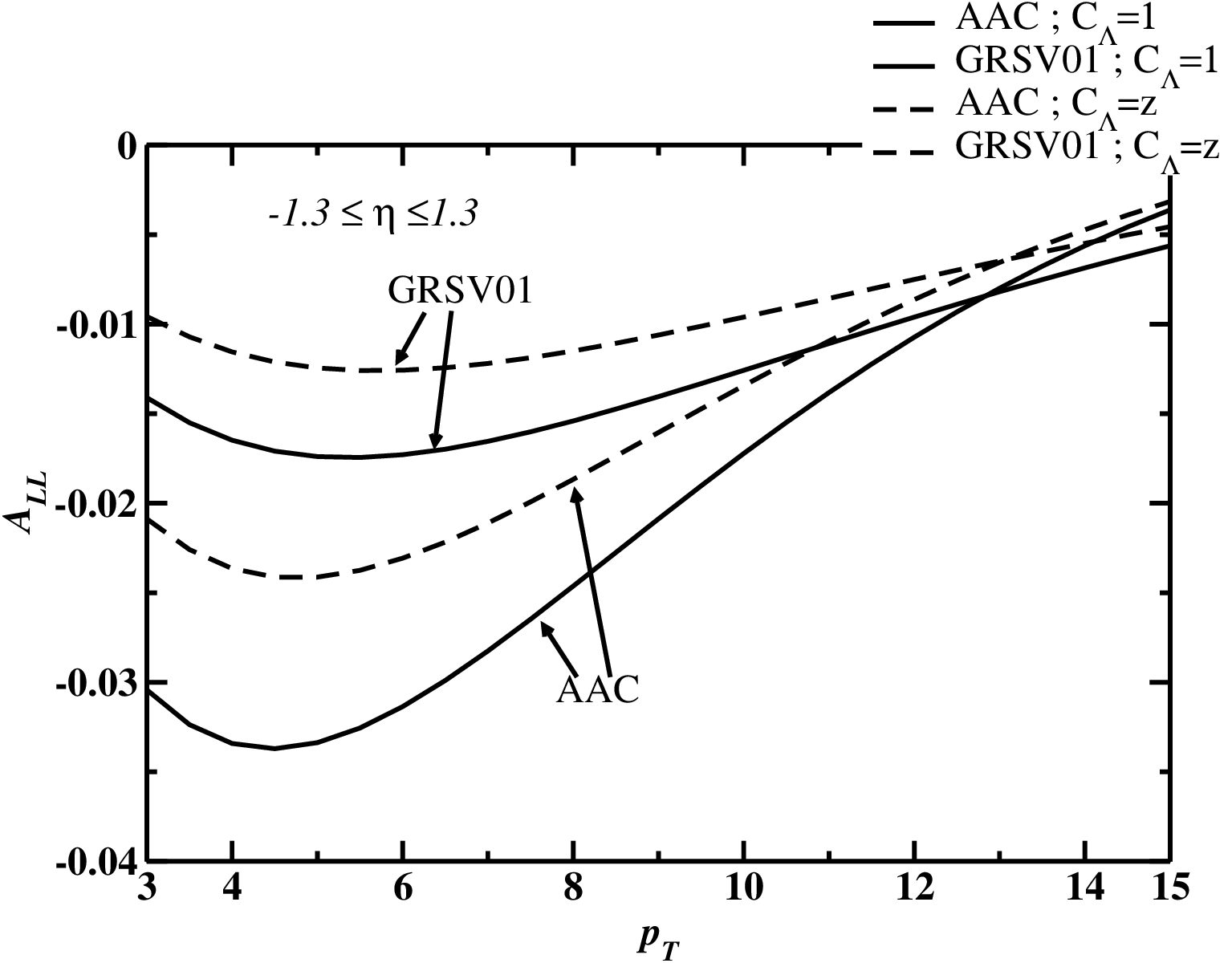}
  \end{center}
  \vspace*{-0.5cm}
}
\caption{$A_{LL}$ as a function of $\eta$ (left panel) and
$\pt$ (right panel) at $\sqrt{s}$=200GeV.} 
\end{figure}
\begin{figure}[htbn]
\hspace*{-1.0cm}
\parbox[b]{0.6\textwidth}
{
  \begin{center}
    \epsfxsize=7.0cm
\epsfbox{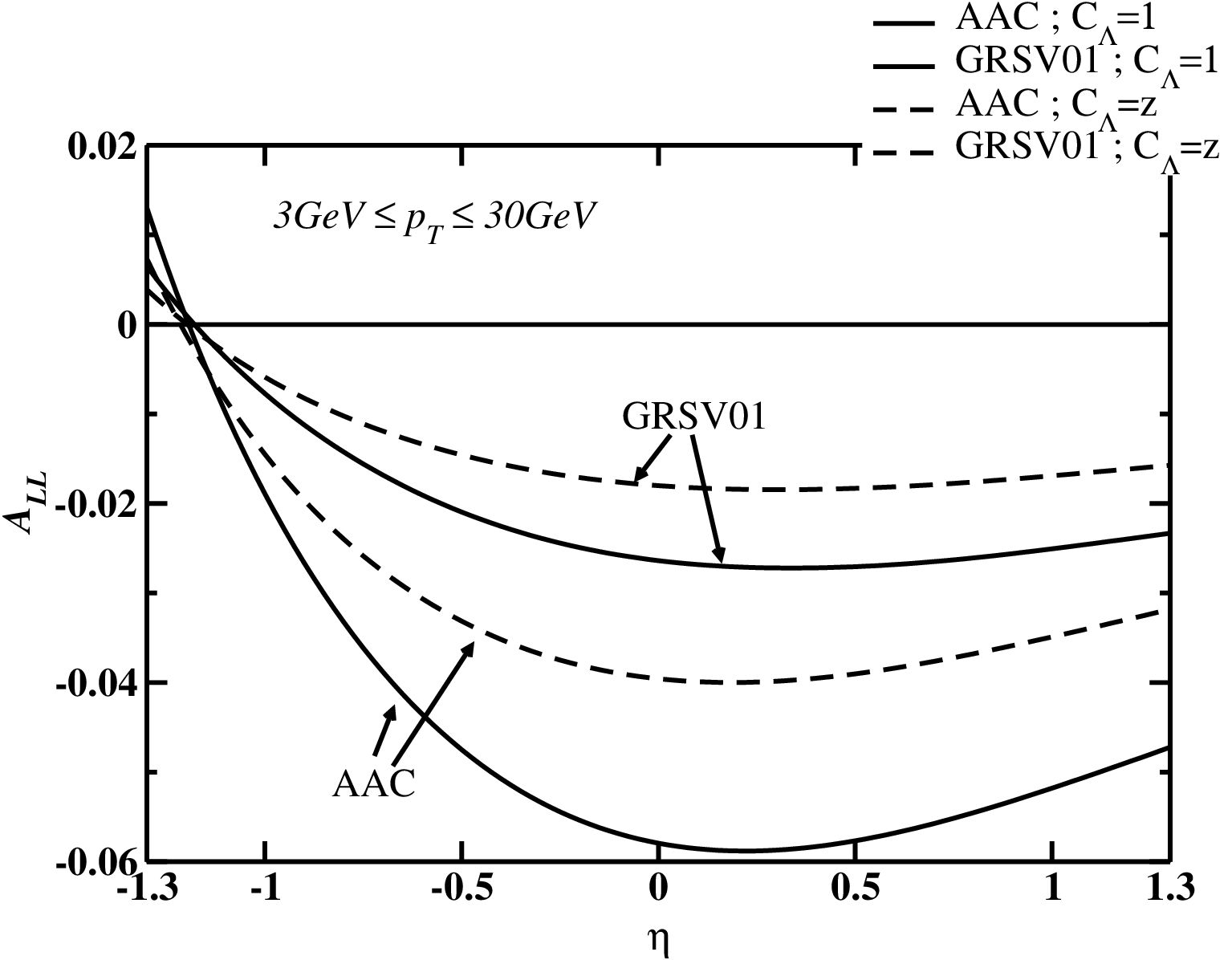}
  \end{center}
  \vspace*{-0.5cm}
} \hfill
\parbox[b]{0.5\textwidth}
{
  \begin{center}
    \epsfxsize=7.0cm
 \epsfbox{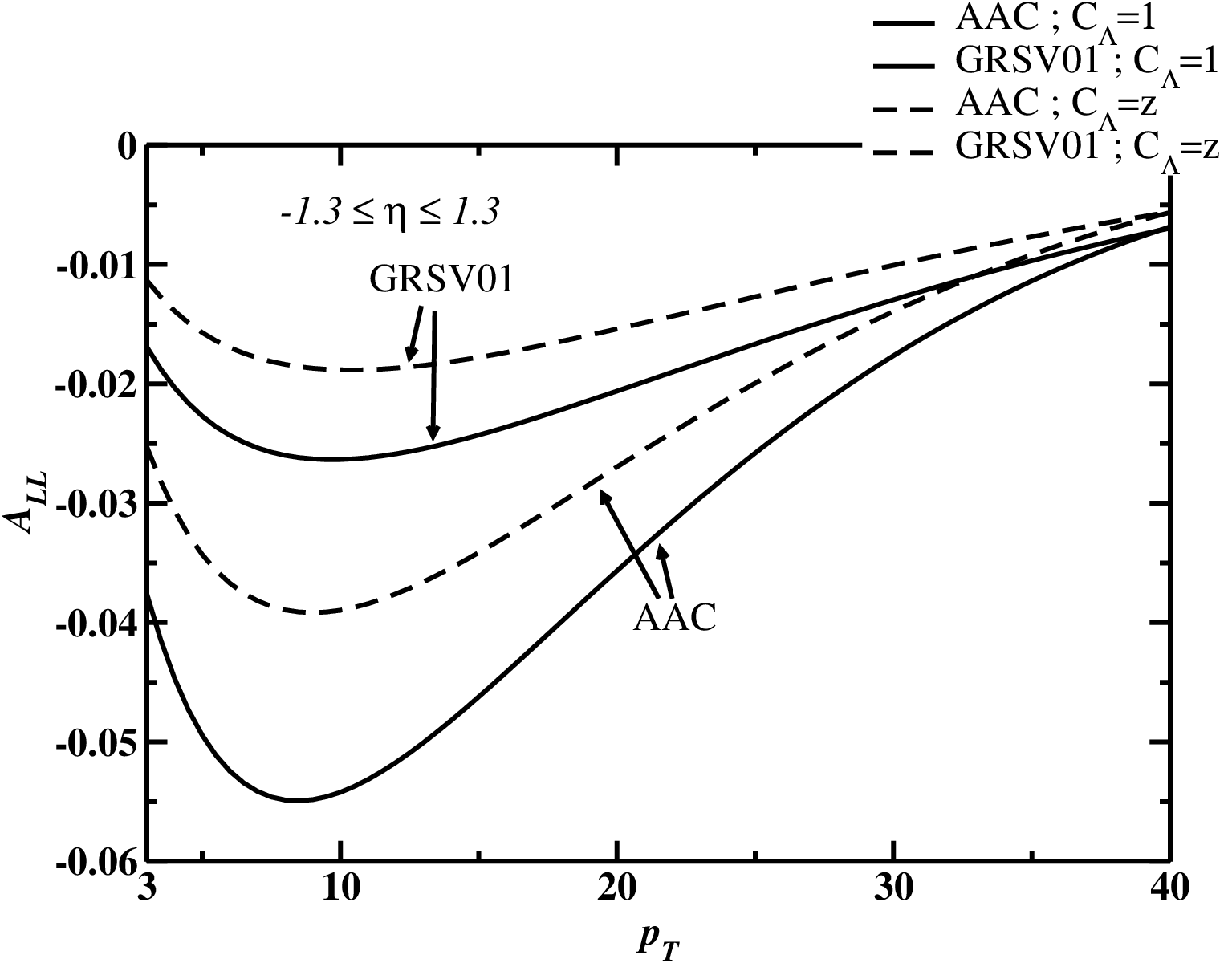}
  \end{center}
  \vspace*{-0.5cm}
}
\caption{$A_{LL}$ as a function of $\eta$ (left panel) and
$\pt$ (right panel) at $\sqrt{s}$=500GeV.} 
\end{figure}


\end{document}